\documentclass[aps,pre,twocolumn]{revtex4}   

\usepackage{amsmath}    
\usepackage{graphicx}   

\def\ba{\begin{eqnarray}}
\def\ea{\end{eqnarray}}
\def\be{\begin{equation}}
\def\ee{\end{equation}}

\begin{document}

\title{On the locus formed by the  maximum heights of 
projectile motion with air resistance
}
\author{H. Hern\'andez-Salda\~na}
\email{hhs@correo.azc.uam.mx}   
\affiliation{Departamento de Ciencias B\'asicas,\\
Universidad Aut\'{o}noma Metropolitana-Azcapotzalco,\\
Av. San Pablo 180, M\'{e}xico 02200 D.F., Mexico.}
\date{\today}

\begin{abstract}
We present an analysis on the geometrical place formed by the set of maxima of the 
trajectories of a projectile launched in a media with linear drag. Such a place, the locus of apexes, 
is written
in term of the Lambert $W$ function in polar coordinates, confirming the special role
played by this function in the problem. In order to characterize the locus, a study of 
its curvature is presented in two parameterizations, in terms of the launch angle 
and in the polar one. The angles of maximum curvature are compared with other 
important angles in the projectile problem. As an addendum, we find that  the synchronous 
curve in this problem is a circle as in the drag-free case.
\end{abstract}

\maketitle

\section{INTRODUCTION}
	An amazing characteristic of some old fashion problems is their endurement. 
The projectile motion is 
one of them. Being one of the main problems used to teach elementary physics, variations and not well 
known facts about  it appear in the physics literature of the XXI century. A search in  
the web \cite{google} or in the {\it Science Citation Index} gives an idea  of this 
fact. 
Some of the recent studies deal with the problem of air resistance in the projectile motion and
its pedagogical character made of it an excellent example to introduce the Lambert $W$ function,
a special function.\cite{Knuth1} 
The Lambert $W$ function is involved in many problems of interest for 
physicist and engineers, from the solution of the jet fuel problem to epidemics\cite{Knuth1} or, 
even, Helium atom eigenfunctions.\cite{Scott}
One of those problems is the solution for 
the range $R$ in the case that the air resistance has the from $\vec{f} = -m b \vec{v}$.\cite{Wang, PackelandYuen}

In this paper we analyze the not well known fact of the geometrical place formed by the maxima of all the projectile 
trajectories at launch angle $\alpha$ and in the presence of a drag force proportional to the velocity, we shall 
denote this locus as ${\cal C}_m(\varepsilon)$. The resulting 
locus becomes a Lambert $W$ function of the polar coordinate $r (\theta)$ departing from the origin.
This problem raises as a natural continuation from the nice fact that  in the drag-free case 
such a locus is an ellipse\cite{Salas,MacMillan,Thomas} with an universal eccentricity 
$e=\sqrt{3}/2$.\cite{Salas}

The paper is organized as follows. In section \ref{Projectilemotion}, the set of maxima for projectile 
trajectories
moving under in the presence of air resistance is presented. In section \ref{solutionLambert} we find
a closed form,
in polar coordinates, to express such a geometrical place, ${\cal C}_m$. In section \ref{param} we present a
numerical calculation of the curvature of ${\cal C}_m$ using the polar angle and the launch 
angle as parameterizations. Additionally, we demonstrate that the synchronous curve is a circle as in the drag-free 
case in section \ref{synchronous}. 
In section \ref{Conclusions} we conclude.

\section{THE PROJECTILE PROBLEM WITH AIR RESISTANCE}\label{Projectilemotion}


Several approximations in order to consider the air resistance exist in the literature, the simplest is 
the linear case. In such a case the force is given by 
\begin{equation}
\vec{F} = -m b \vec{v} -m g \hat{\j}, 
\label{EqNewton}
\end{equation}
where $m$ is the mass of the projectile and $b$ is the drag coefficient. The units of $b$  are $s^{-1}$. 
The velocity components are 
labeled as $\vec{v} = u \hat{\i} + w \hat{\j}$, with $u = dx/dt$ and $w=dy/dt$.
The solutions for the position and velocity are obtained trough direct integration of 
Eq.(\ref{EqNewton}) yielding 
\be
\begin{array}{rcl}
x(t) & = & \frac{u_0}{b} \left[ 1 -\exp(-b t)\right], 
\label{timesolx}
\end{array}
\ee
\be
\begin{array}{rcl}
y(t) & = & \frac{w_0+g/b}{b}\left[ 1- \exp(-b t)\right] -g t/b,
\label{timesol}
\end{array}
\ee
for the coordinates, and 
\begin{eqnarray}
u(t) & = & u_0 \exp(-b t), \\
w(t) & = &  (w_0+g/b) \exp(-b t) - g/b,
\end{eqnarray}
for the speeds. We used 
the initial conditions $x(0)=y(0)=0$ and $u_0 = V_0 \cos \alpha$ and $w_0 = V_0 \sin \alpha$.
Noticing that the terminal speed is $g/b$ in the $y$ axis.

\begin{figure}[tb]
\includegraphics[width=\columnwidth]{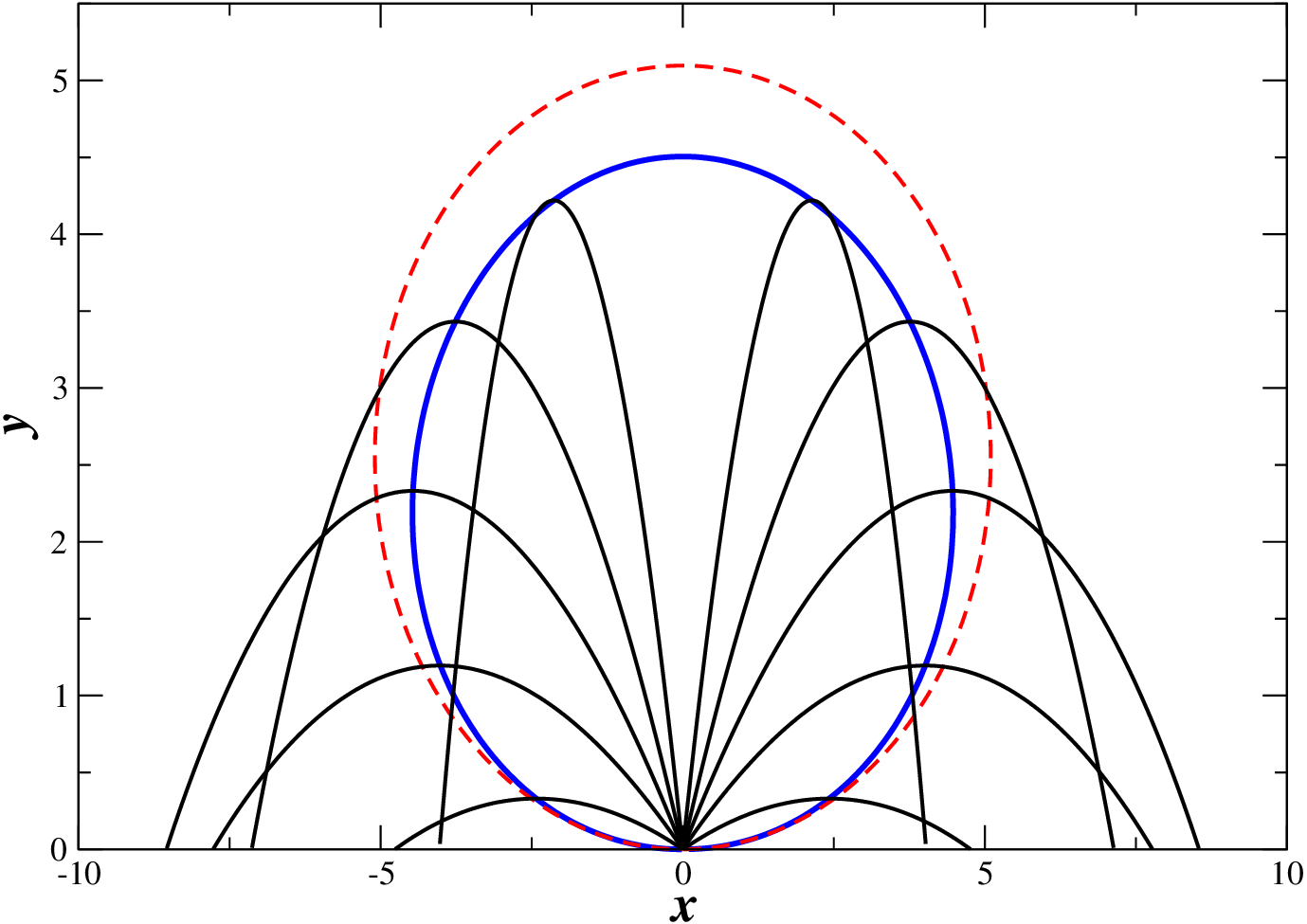}
\caption{(color online) Locus ${\cal C}_m(\varepsilon)$ formed by the apexes of all the projectile 
trajectories (continuous line in blue) given by Eqs. (\ref{xmeqno}) and (\ref{ymeqno}) in rectangular 
coordinates or by Eq. (\ref{W1}), the last one express ${\cal C}_m$ in polar coordinates and in term of the Lambert $W$ function.  
The dashed red line is the ellipse of eccentricity $e=\sqrt{3}/2$ which represents the 
drag-free case, i.e. ${\cal C}_m(0)$.  
The parameters are $V_0= 10$  and $\varepsilon = 0.1$.}
\label{Fig:GeometricPlace}
\end{figure}

For the same initial speed $V_0$ these solutions are function of the launch angle $\alpha$ and the locus
formed  by the apexes is obtained  if time is eliminated between the solutions in 
time in Eqs. (\ref{timesolx}) and (\ref{timesol}), giving the equation
\begin{equation}
y(x)  =  \frac{w_0+g/b}{u_0} x -\frac{g}{b^2} \ln \left( 1- \frac{b x}{u_0}\right),
\end{equation}
and considering the value at the maximum, via $dy/dx= 0$. 
The corresponding solution is
\begin{equation}
\frac{x_m}{\rho} =\cos \alpha \sin \alpha \frac{1}{1+\varepsilon \sin \alpha},
\label{xmeqno}
\end{equation}

\begin{equation}
\frac{ \varepsilon^2 y_m}{\rho} = \varepsilon \sin \alpha - \ln(1+\varepsilon \sin \alpha),
\label{ymeqno}
\end{equation}
where we introduce the dimensionless perturbative parameter $\varepsilon \equiv b V_0 / g$, the dimensionless
length $\rho = V_0^2/g$,
and noticing that $\frac{b^2 }{g}$ can be expressed as $\frac{\varepsilon^2 }{\rho}$. 
An alternative procedure consists in set the 
derivative $dy/dt$ to zero to obtain the time of flight to the apex of the trajectory and, evaluate the 
coordinates at that time. 
The points $(x_m,y_m)$ conform the locus of apexes  ${\cal C}_m(\varepsilon)$
for all parabolic trajectories  as a function of the launch angle $\alpha$. 
In Fig. \ref{Fig:GeometricPlace} we plot ${\cal C}_m(\varepsilon)$ described by 
Eqs. (\ref{xmeqno}) and (\ref{ymeqno}), for the drag-free case (in dashed red line) and for $\varepsilon = 0.1$ 
in continuous blue line. Several  projectile trajectories  are plotted in thin black lines.
The locus of apexes ${\cal C}_m(\varepsilon)$ defined by $(x_m,y_m)$ of Eqs. (\ref{xmeqno}) and (\ref{ymeqno}) is described parametrically by the launch angle $\alpha$ and it changes for different values of 
$\varepsilon \equiv b V_0 / g$. In the next section we shall find a description of ${\cal C}_m(\varepsilon)$ 
in terms of polar coordinates and in a closed form using the Lambert $W$ function.

\section{The locus ${\cal C}_m$ as a Lambert $W$ function}\label{solutionLambert}

In order to obtain an analytical closed form  
of the locus we change the variables to polar ones, i.e., $x_m = r_m \cos \theta_m$ and $y_m = r_m \sin \theta_m$.
The selection of a description departing from that origin instead of the center or the focus  of 
the ellipse is because the resulting geometrical place is no longer symmetric and the 
only invariant point is just the launching origin.  We substitute the polar forms of $x_m$ and 
$y_m$ into equations (\ref{xmeqno}) and (\ref{ymeqno}) and
rearranging terms it must be expressed as
\begin{eqnarray}
\begin{array}{l}
\frac{r_m(\theta_m)}{\rho} \cos \theta_m
\exp\left(-\varepsilon^2 \sin \theta_m  \frac{r_m(\theta_m)}{\rho}\right)  \\
\ \ \ \  \ \  \ \ \ = \cos \alpha \sin \alpha \exp(-\varepsilon \sin \alpha).
\end{array}
\label{ymeqno2}
\end{eqnarray}
The lhs depends on $r_m$ and $\theta_m$ meanwhile the rhs depends on $\alpha$, however
the last angle is a function of $\theta_m$ and reads as
\begin{equation}
\tan \theta_m =\frac{1}{\varepsilon^2} 
\frac{\left( \varepsilon \sin \alpha - \ln(1+\varepsilon \sin \alpha) \right)}
{\cos \alpha \sin \alpha \frac{1}{1+\varepsilon \sin \alpha}},
\label{angleEq}
\end{equation}
by making $\tan \theta_m = y_m/x_m$ from Eqs. (\ref{xmeqno}) and (\ref{ymeqno}).

In order to obtain $ \tilde{r} (\theta_m) \equiv r_m(\theta_m)/\rho$ we set 
\begin{equation}
f(\alpha (\theta_m)) \equiv \cos \alpha \sin \alpha \exp(-\varepsilon \sin \alpha),
\label{fdtheta}
\end{equation}
since Eq. (\ref{angleEq}) allows us to have, implicitly, $\alpha (\theta_m)$. 
We shall return to this point later.
Hence, we can write Eq. (\ref{ymeqno2}) as 
\begin{equation}
-\varepsilon^2 \sin \theta_m \ \tilde{r}(\theta_m) 
\exp\left(-\varepsilon^2 \sin \theta_m \ \tilde{r}(\theta_m)\right)
= -\varepsilon^2 \tan \theta_m \ f(\alpha).
\label{pre1}
\end{equation}
Where we multiplied both sides of Eq. (\ref{ymeqno2}) by $-\varepsilon^2 \sin \theta_m$. 
Setting $z = -\varepsilon^2 \tan \theta_m \ f(\alpha) $ and 
$W(z) = -\varepsilon^2 \sin \theta_m \ \tilde{r}(\theta_m)$
in Eq. (\ref{pre1}), it shall have the familiar Lambert $W$ function form, 
$z= W(z) \exp(W(z))$, from which 
we can obtain $\tilde{r}$ as
\begin{equation}
\tilde{r}(\theta_m) = -\frac{1}{\varepsilon^2 \sin \theta_m} W(-\varepsilon^2 \tan \theta_m 
\ f(\alpha)).
\label{W1}
\end{equation}

It is important to note that the argument of the Lambert function in this equation is negative for all the 
values $\varepsilon > 0$. $W(x)$ remains real in the range  $x\in [-1/e,0)$ and have the branches
denoted by $0$ and $-1$.\cite{Knuth1} We select the principal branch, $0$, since it is the bounded 
one, however, for values of $\varepsilon > 1.1$ there is a precision problem since the required
argument values are near to $-1/e \equiv -\exp(-1)$. 
It is important to stress that in Eq. (\ref{W1}) the independent variable is the angle $\theta$ and, 
it constitutes the parameterization of the curve ${\cal C}_m$. 

The polar expression of ${\cal C}_m$ can also be written in terms of the tree function $T(z) = -W(-z)$, giving
\begin{equation}
\tilde{r}(\theta_m) = \frac{1}{\varepsilon^2 \sin \theta_m} T(\varepsilon^2 \tan \theta_m 
f(\alpha(\theta_m))).
\label{W2}
\end{equation}
We recover the drag-free result
\be
\tilde{r} = 2\frac{\sin \theta_m }{1+3\sin^2 \theta_m}
\label{ellipse0}
\ee
when $\varepsilon \to 0$. An explanation of this unfamiliar form of an ellipse is given in appendix
\ref{appendix1} followed by a discussion about the $\varepsilon \to 0$ limit of expression (\ref{W1}) 
in appendix \ref{appendix2}.

Formula (\ref{W1}) exhibits the deep relationship between the Lambert $W$ function and the linear drag 
force projectile problem, since not only  the range is given as this 
function \cite{Wang,PackelandYuen}. The problem open the opportunity to study the W function in polar 
coordinates, that, almost in the review of referencei~ \cite{Knuth1}, is absent. Even when it is possible to 
write  the locus in terms of $y_m(x_m)$ this form does not shows the formal elegance of relation (\ref{W1}).

\begin{figure}[tb]
\includegraphics[width=\columnwidth]{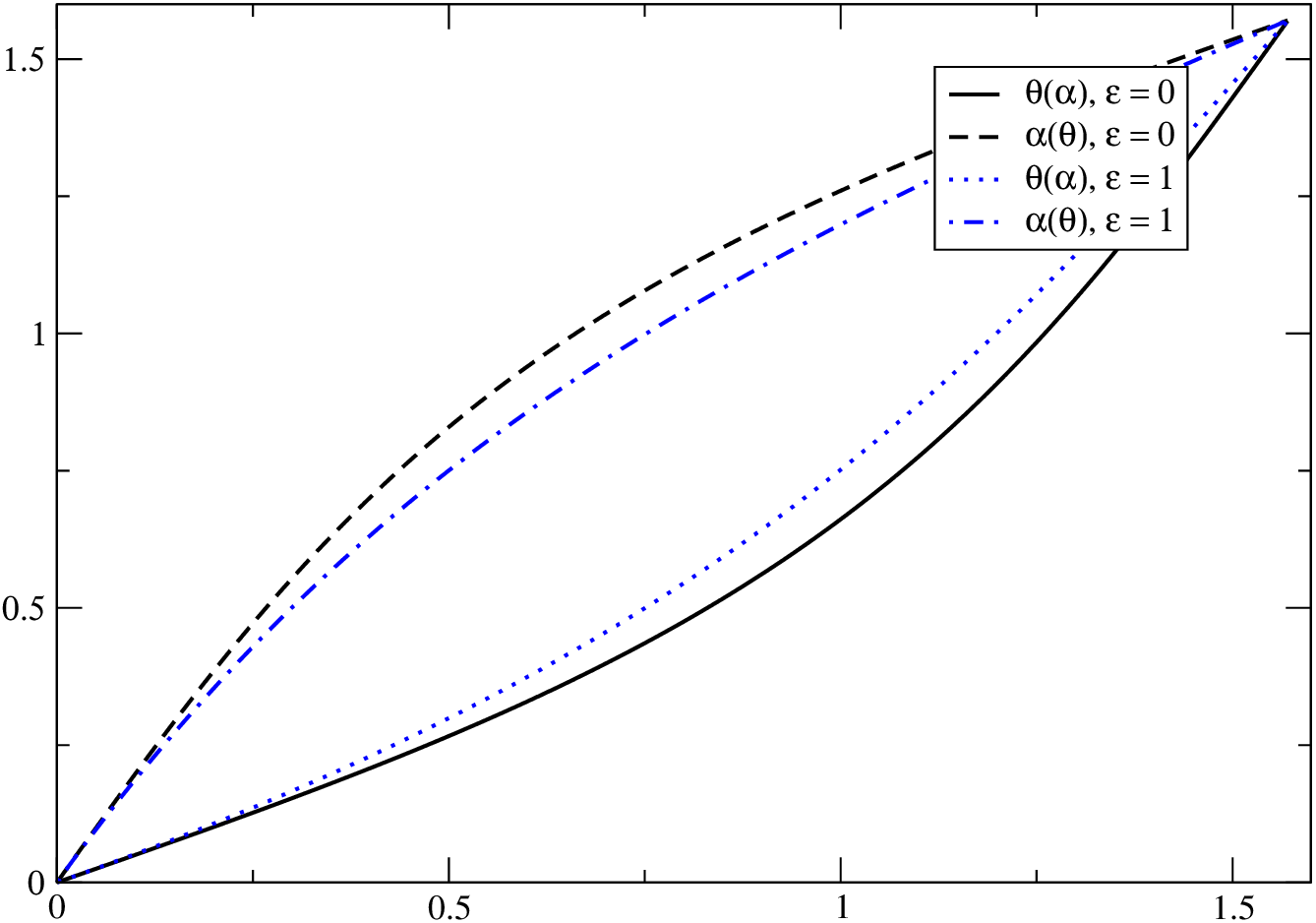}
\caption{(color online) The angle $\theta$ as a function of the launch  angle $\alpha$ for two different
values of $\varepsilon$ and their inverses.}
\label{Fig:inverse}
\end{figure}

Now we return to equation (\ref{angleEq}) since we need to solve explicitly it in order 
to have the function $\alpha(\theta_m)$. This task is not trivial
since even when we approximate the rhs in expression  (\ref{angleEq}) up to first order 
in $\varepsilon$,
\begin{equation}
\tan \theta_m = \frac{\tan \alpha }{2}(1+\frac{\varepsilon}{3}\sin \alpha),
\end{equation}
the inversion is not easy. A way to do the inversion is to expand in a Taylor series the rhs and 
then invert the series term by term.\cite{Arfken}
Using {\it Mathematica} to perform this procedure up to ${\cal O}(18)$, we obtain  as a result 
\ba
\begin{array}{ll}
\alpha(\theta) \approx & \arctan(2 \tan \theta ) -\frac{1}{3} \varepsilon (2 \tan \theta)^2
      + \frac{2}{9} \varepsilon^2 (2 \tan \theta )^3 \\
 &          + \frac{1}{54} (27 \varepsilon - 10 \varepsilon^3) (2 \tan \theta ))^4 + \cdots
\label{expantion}
\end{array}
\ea
The $\varepsilon$-independent terms had been resumated
to yield $\arctan(2\tan \theta)$. However, the series does not converge for values in the argument 
larger than $1$. The reason is the small convergence ratio for the Taylor expansion of $\arcsin(\cdot)$. 

\begin{figure}[tb]
\includegraphics[width=\columnwidth]{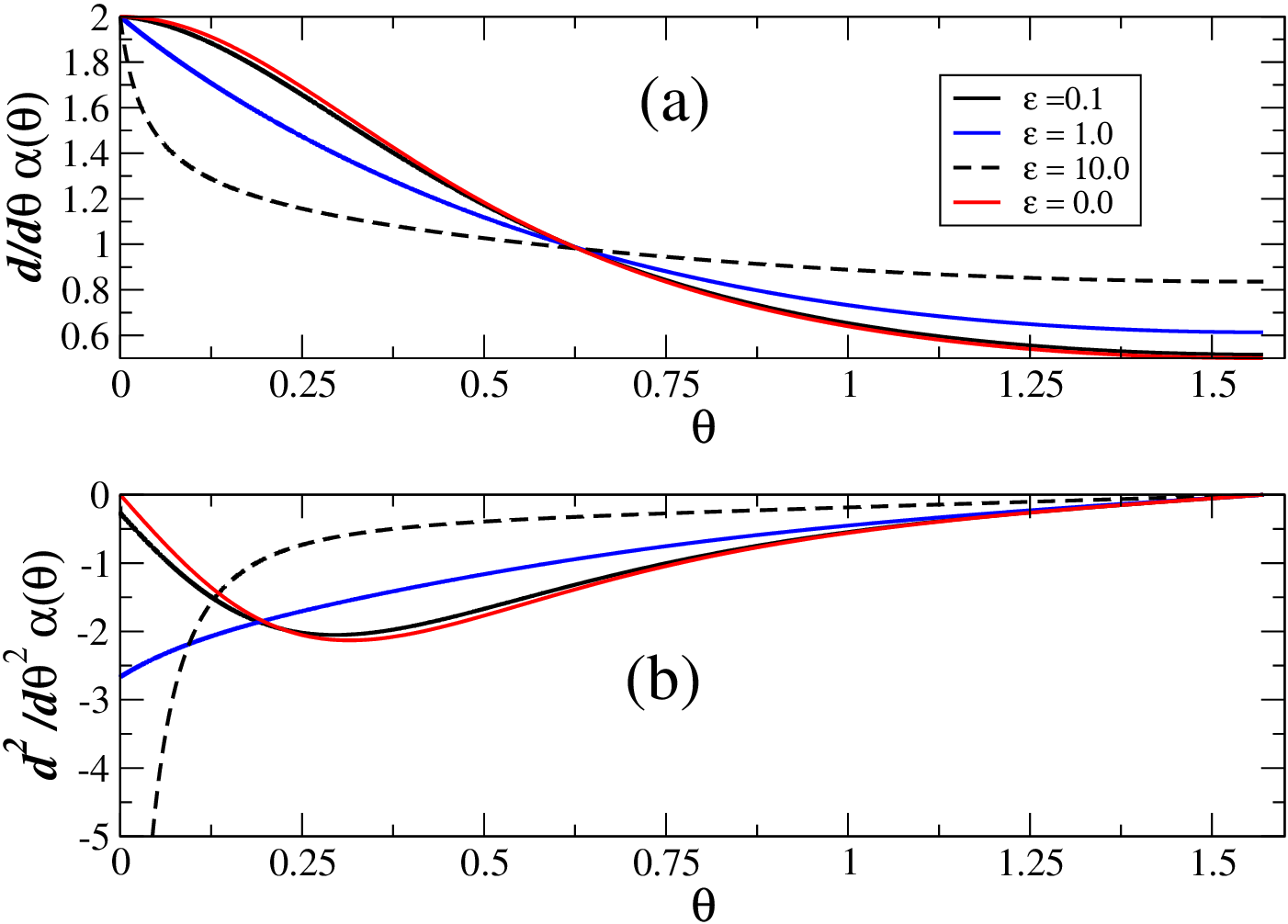}
\caption{(color online) (a) First and (b) second derivatives of $\alpha$ as function of $\theta$
for various values of parameter $\varepsilon$. Note that major changes occur for $\theta < \pi/4$.
}
\label{Fig:invderiv}
\end{figure}

An easier way to perform the inversion  is to evaluate $\theta_m(\alpha)$ using Eq. (\ref{angleEq}) and plot the 
points $(\theta_m(\alpha),\alpha)$, the result is shown in figure \ref{Fig:inverse}. 
The result is in agreement with the plot of Eq. (\ref{expantion}) up to its convergence ratio and it is not shown. 
Notice that this method is exact in the sense that we can obtain as many pair of numbers as we need, a function
is, finally, a relation one to one between two sets of real numbers.
Another result 
is to obtain the derivative $d \alpha/d\theta$, since it shall be needed in the following sections.
To this end, we note that both functions increase monotonically and their derivatives are not 
zero, except at the interval end. Hence, we can use the inverse function theorem in order to obtain 
\be
\frac{d \alpha}{d\theta} = \frac{1}{\frac{d \theta}{d\alpha}}.
\ee
The result is shown in Fig. \ref{Fig:invderiv}(a) as well as the second derivative in 
Fig. \ref{Fig:invderiv}(b). The second derivative is calculated using an approximation to 
the slope to the function previously calculated and using $10000$ points in the interval
$[0,\pi/2]$. A smaller number of points could be considered.

\section{The curvature of ${\cal C}_m$.}\label{param}
\subsection{Polar angle parameterization.}

In the drag-free situation,  ${\cal C}_m$ is an ellipse and its description is 
well know, however, in the presence of linear drag this is not the case. 
We do not expect that the locus could be a conic section and henceforth we need to 
characterize it. It is usual to consider curvature, radius of curvature or the length of 
arc in order to characterize a locus. In the present case  we consider the curvature  of  ${\cal C}_m$ 
in both parameterizations, first with the polar angle $\theta_m$
and secondly with the launch angle $\alpha$. We left the calculus of the length of arc 
to a posterior work, since the calculations became increasingly complex and the goal of the present section 
is to start the understanding of  ${\cal C}_m$ and to illustrate the way it can be done using 
the Lambert  W function. Here and in the rest  of the section we drop, for clearness, the subindex 
$m$ in $\tilde{r}_m$ and $\theta_m$.

The corresponding formula for the curvature $K$ for polar coordinates is \cite{Gray}
\begin{equation}
K =\frac{1}{\rho} \frac{ \tilde{r}^2+2\tilde{r}_{\theta}^2 - \tilde{r} \tilde{r}_{\theta  \theta}}
{(\tilde{r}^2+\tilde{r}_{\theta}^2)^{3/2}},
\label{Kappa}
\end{equation}
in order to use Eq. (\ref{W1}). Here the subindex $\theta$ corresponds a derivative respect to that 
variable. 

\begin{figure}[tb]
\includegraphics[width=\columnwidth]{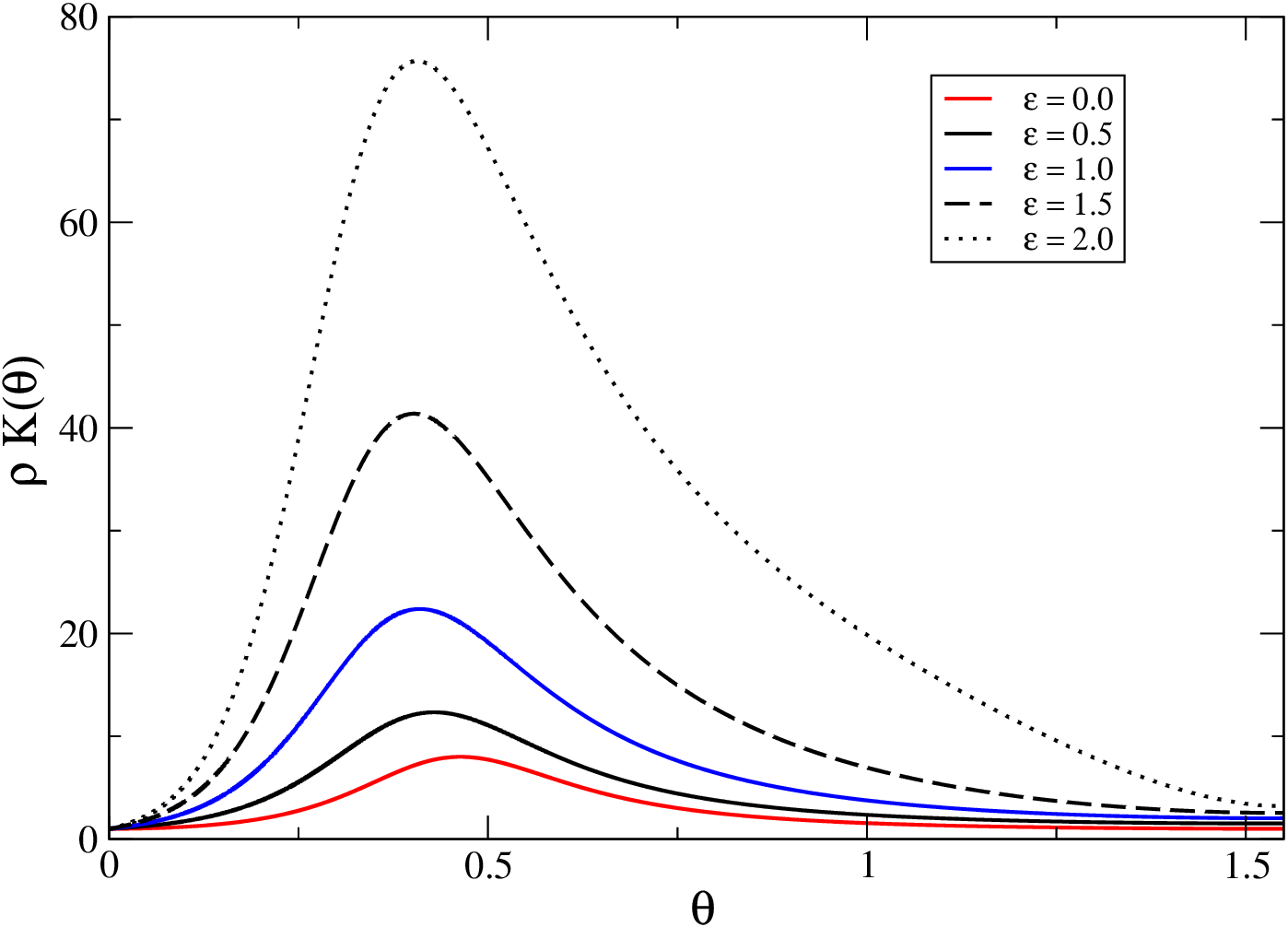}
\caption{(color online) Curvature of ${\cal C}_m(\varepsilon)$ using the polar angle as 
the parameter from Eq. (\ref{Kappa}). In red line appears the corresponding result for the ellipse, 
Eq. (\ref{Kappa0}). The maximum happens at $\theta = 1/2 \arctan(4/3)$, different from the 
value of maximum range $\alpha= \pi/4$.
}
\label{Fig:Kpolar}
\end{figure}

A direct calculation on the drag-free $\tilde{r}(\theta)$ of Eq.(\ref{ellipse0}) 
yields to 
\be
K_0 = \frac{1}{2\sqrt{2}\rho}
\frac{(5 - 3 \cos 2 \theta )^6}{(1 + 3\sin^2 \theta)^3 
(47 - 60\cos 2\theta + 21\cos 4\theta )^{3/2}},
\label{Kappa0}
\ee
which  have a maximum at $\theta = 1/2 \arctan(4/3) \approx 0.464$. A graph of
this results appears in figure \ref{Fig:Kpolar}(red line).  Note that this value 
is different from that  we obtain if we evaluate $\theta( \alpha = \pi/4) = 1/2$, the 
launch angle of  maximum range. 
The maximum curvature happens at a smaller angle than the angle of maximun range.
It is interesting to note that the angle  
$2 \theta = \arctan(4/3)$ corresponds to a triangle which sides fulfill 
the relation $3^2+4^2=5^2$, a Pythagoras' triple.

Using the numerical results for $\alpha(\theta)$ from the previous section, it is 
possible to carry on the calculation of $K$ (see figure \ref{Fig:Kpolar}) performing
the derivatives of $\tilde{r}$ from Eq. (\ref{W1}) in a direct
form and evaluating numerically the required  values of $\alpha$ and its derivatives. 
For the required derivatives of $W(z)$ we used the expressions \cite{Knuth1} 
\be
\frac{d}{d x}W(x) = \frac{W(x)}{x (1+ W(x))},
\ee
and 
\be
\frac{d^2 W(x)}{dx^2} = \frac{ -\exp(-2 W(x))(W(x)+2) } {(1+ W(x))^3}.
\ee
Using this method, we obtained  good results
for values of $\varepsilon$ up to  $\approx 1$ but we require to calculate arguments 
of the Lambert $W$ function near the limit $z = -1/e$ for larger values of $\varepsilon$.
The reliability  of our numerical result was done comparing the first and second 
derivatives of $r(\theta)$ with those corresponding to the ellipse.

As can be seen in figure \ref{Fig:Kpolar}, $K$ present a maximum in all  the cases
which can be calculated as well. We left to an ulterior work the analysis of the 
maxima distribution as a function of the pertubative parameter, that is not the 
case for the curvature with $\alpha$ parameterization as we shall see in the next section.

\subsection{Launch angle parameterization.}
For the launch  angle parameterization of ${\cal C}_m$ we shall use expression \cite{Gray}
\begin{equation}
\kappa =\frac{ x' y''- y' x''}{(x'^2+y'^2)^{3/2}},
\label{kappa}
\end{equation}
for calculate the curvature from the rectangular form of equations 
(\ref{xmeqno}) and (\ref{ymeqno}), where
the primes denote derivative respect the parameterization variable, $\alpha$ in this case.

A direct calculation yields
\ba
\begin{array}{lll}
\kappa &=&\frac{\sqrt{2}}{\rho} 
\frac{ {\cal P}_1(\alpha)} {(\sqrt{{\cal P}_2(\alpha)})^3} \times (1+\varepsilon \sin \alpha)^2. \\
\label{kappa_e}
\end{array}
\ea
With
\ba
\begin{array}{ll}
{\cal P}_1 (\alpha)= &
16+6\varepsilon^2-8\varepsilon^2\cos 2\alpha+ 2\varepsilon^2 \cos 4\alpha  \\
 &+ 30\varepsilon \sin \alpha -\varepsilon \sin 3\alpha + \varepsilon \sin 5\alpha,
\end{array}
\ea
and
\ba
\begin{array}{ll}
{\cal P}_2 (\alpha)= & 5 + 3\cos 4\alpha  +3\varepsilon^2 -4\varepsilon^2 \cos 2 \alpha  
+ \\ 
 & \varepsilon^2 \cos 4\alpha+10\varepsilon \sin \alpha - 5\varepsilon\sin 3\alpha + \\
 & \varepsilon \sin 5\alpha.
\end{array}
\ea

\begin{figure}[tb]
\includegraphics[width=\columnwidth]{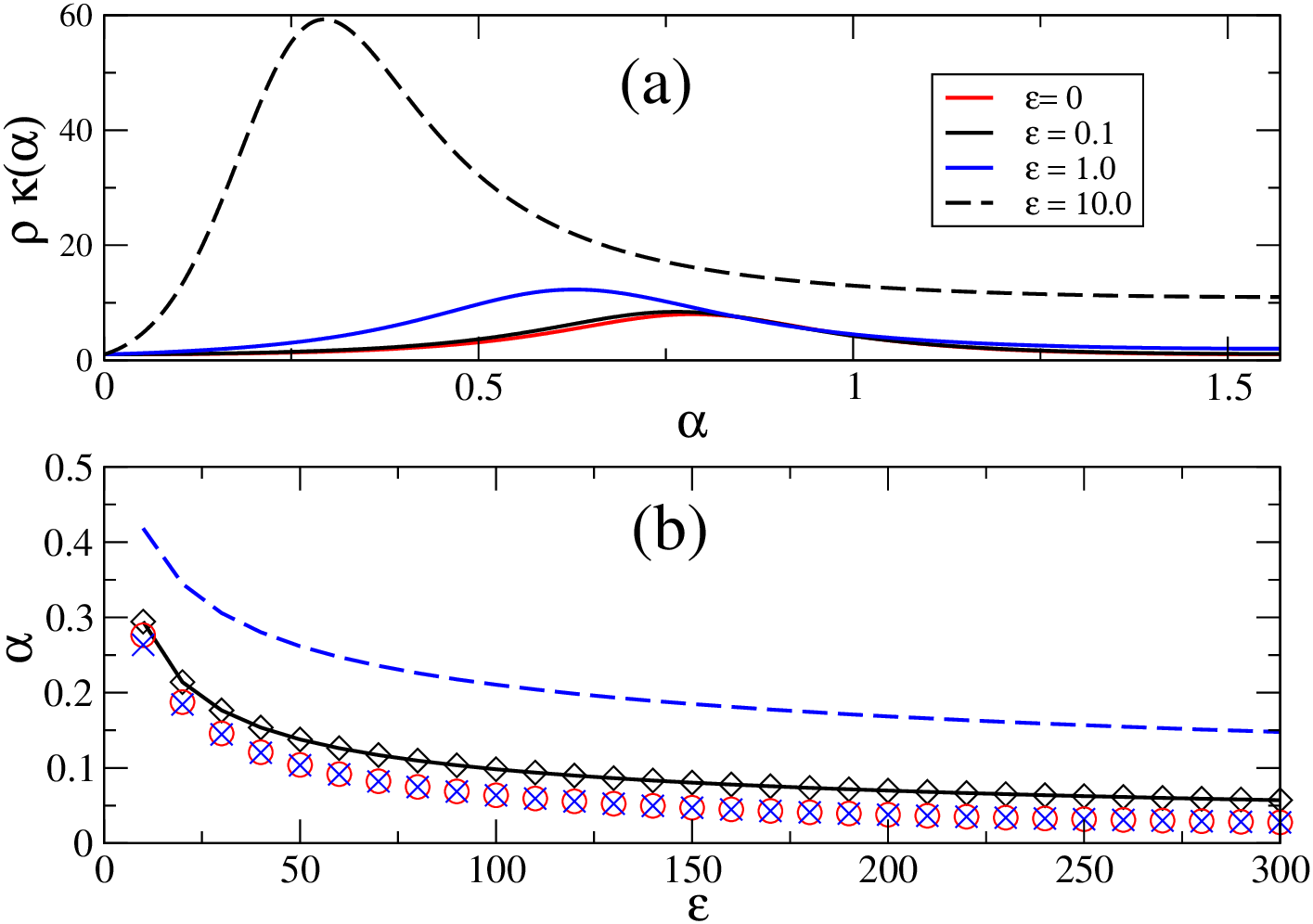}
\caption{(color online) (a) Curvature as a function of the launch angle $\alpha$ for increasing 
values of parameter $\varepsilon$ according to equation (\ref{kappa_e}). 
The values of $\varepsilon$ are indicated in the inset. 
(b) Several important angles as a function of dimensionless parameter $\varepsilon$ are plotted. 
In lines and diamonds
appears the angle, $\alpha^*$, at which the curvature is maximum. In red circles the angle at which 
the range is maximum according to exact solution, Eq. (\ref{exactangle}), and in blue crosses the 
same angle according to Eq. (\ref{asympangle})
(see text for discussion). In a dashed blue line  the angle at which skewness is maximum is plotted.
}
\label{Fig:angle}
\end{figure}

In the limit $\varepsilon \to 0$, we recover the drag-free curvature 
\be
\kappa_0 = \frac{16 \sqrt{2}}{\rho} \frac{1}{(5+3 \cos 4\alpha )^{3/2}},
\label{kappa_0}
\ee
which have a maximum at $\alpha = \pi/4$ in the interval $\alpha \in [0,\pi/2]$,  as expected. 
A plot of $\rho \kappa(\alpha)$ for several values of $\varepsilon$ beginning at zero and 
ending at $\varepsilon = 10 $ 
appears in Figure \ref{Fig:angle}(a). In red appears the drag-free case. 
Note that both extremal values increase for increasing $\varepsilon$ value as 
$\rho \kappa(0) \approx (16+6\varepsilon-6\varepsilon^2)$ and $\rho \kappa(\pi/2) \approx \varepsilon $. 
Notice that for small  $\varepsilon$, the $\kappa$ crosses the drag-free curvature $\kappa_0$.

The angles $\alpha^*$ at which $\kappa$ attain their maxima are obtained in the usual way and requires to 
solve, numerical or graphically, the equation
\ba
\begin{array}{l}
 3 (1 + \varepsilon \sin \alpha^*) \sin 2\alpha^* \times
{\cal Q}_1(\alpha^*) {\cal Q}_2(\alpha^*) + \\
\varepsilon \cos \alpha^*  {\cal Q}_3(\alpha^*) {\cal Q}_4(\alpha^*) = 0,
\end{array}
\ea
with
\ba
\begin{array}{ll}
{\cal Q}_1 = &
4\ (3 + \varepsilon^2) \cos 2 \alpha^* + \varepsilon (-4  \varepsilon - 15\sin \alpha^* + \\
 & 5 \sin 3\alpha^*);
\end{array}
\ea
\ba
\begin{array}{ll}
{\cal Q}_2 = &
16 + 6 \varepsilon^2 - 8 \varepsilon^2  \cos 2 \alpha^* + 2 \varepsilon^2 \cos 4 \alpha^* + \\
 & 30 \varepsilon \sin \alpha^* - \varepsilon \sin 3 \alpha^* + \varepsilon \sin 5 \alpha^*;
\end{array}
\ea
\ba
\begin{array}{ll}
{\cal Q}_3 = &
5 + 3 \cos 4 \alpha^*+ 3\varepsilon^2 - 4 \varepsilon^2 \cos 2 \alpha^* + \\
 & \varepsilon^2 \cos 4 \alpha^* + 10\varepsilon \sin \alpha^* - \\
 & 5\varepsilon\sin 3\alpha^* + \varepsilon \sin 5 \alpha^*;
\end{array}
\ea
and
\ba
\begin{array}{ll}
{\cal Q}_4 = &
70 + 36 \varepsilon^2 - 16 (1 + 3 \varepsilon^2) \cos 2 \alpha^* + \\
 & 2 (5 + 6\ \varepsilon^2) \cos 4 \alpha^* +
154 \varepsilon \sin \alpha^* - 31 \varepsilon \sin 3\ \alpha^* \\
 & + 7 \varepsilon \sin 5 \alpha^*.
\end{array}
\ea

In Figure \ref{Fig:angle}(b) the calculated values  of $\alpha^*$ as a function of $\varepsilon$ appear. 
This angle is 
between the optimal angle for maximum range (red circles and blue crosses) and  the  angle for the greatest 
forward skew (dashed line).\cite{Steward2006}
\be
\alpha_{skew} = \arcsin\left[\frac{1}{3\varepsilon}\left((D_+/2)^{1/3} + (D_-/2)^{1/3} - 2\right)\right],
\ee
where $D_{\pm} = \pm 3\varepsilon \sqrt{3(27\varepsilon^2 - 32)}+ 27\varepsilon^2-16$, valid for 
$\varepsilon > \sqrt{32/27}$.\cite{Comment}
In Figure \ref{Fig:angle}(b) the optimal angles are drawn, in red circles the 
exact result in terms of Lambert $W$ function\cite{PackelandYuen,Steward2005a}
\be
\alpha_{max,S} = \arcsin\left[ \frac{\varepsilon}{\exp\left(W \left(\frac{\varepsilon^2-1}{e}\right) + 1 \right)-1}
\right],
\label{exactangle}
\ee 
and the approximated result\cite{Wang}
\be
\alpha_{max,W} = \frac{ W(\varepsilon^2/e)}{\varepsilon}.
\label{asympangle}
\ee
Both expressions are equivalent for large $\varepsilon$ but differ at small $\varepsilon$, as expected.

\begin{figure}[tb]
\includegraphics[width=\columnwidth]{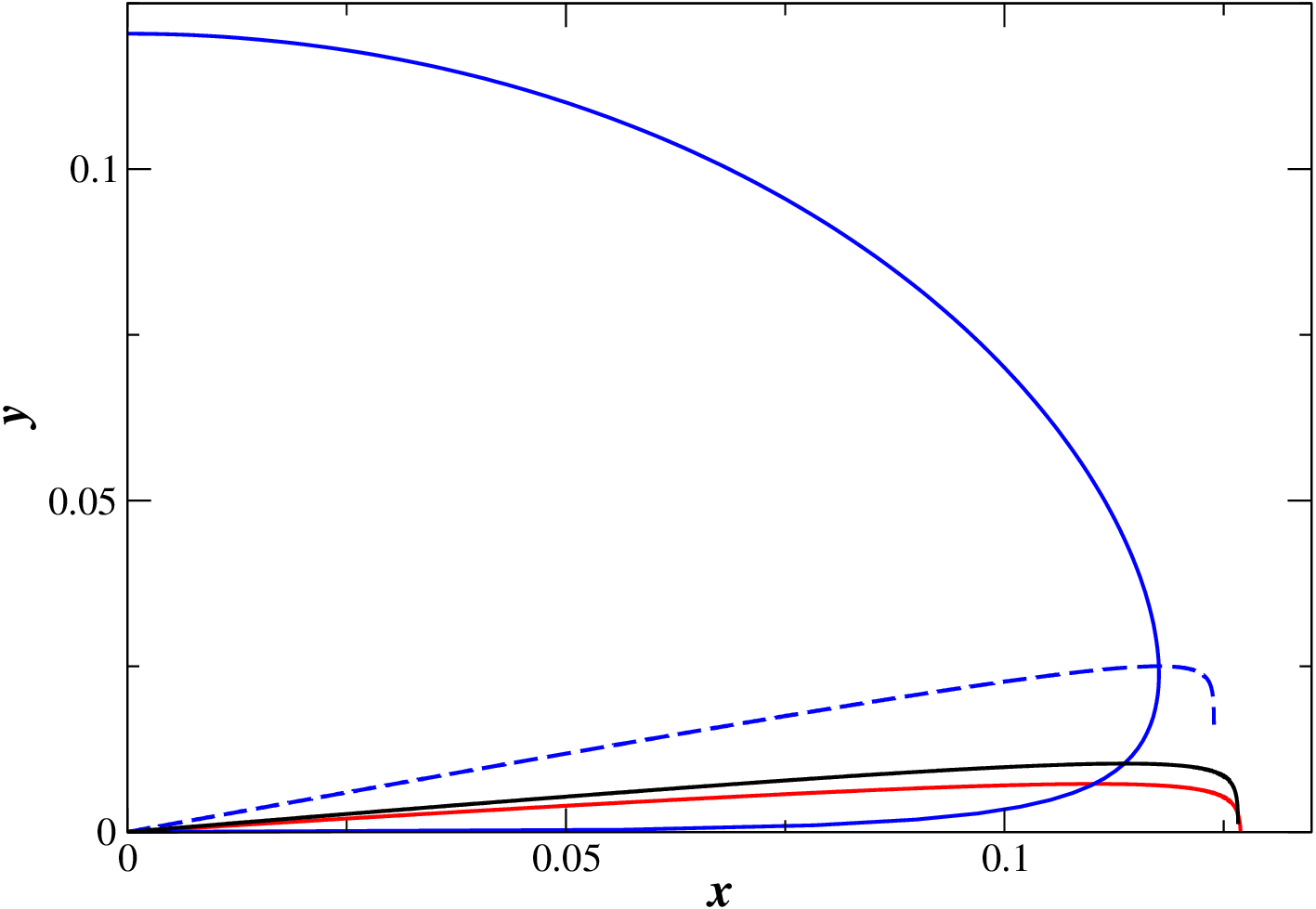}
\caption{(color online) Orbits launched at different angles and the corresponding geometrical place
${\cal C}_m(\varepsilon)$ for $\varepsilon=80$. See text for explanation.}
\label{Fig:LargeEps}
\end{figure}

Meanwhile the difference between these angles at small pertubative parameter is unimportant, 
at large $\varepsilon$ the behavior of the corresponding trajectories is different. One reason is the 
large asymmetry in the locus formed by the set of apexes. In Fig. \ref{Fig:LargeEps}(a)
we plotted ${\cal C}_m(\varepsilon)$ and the corresponding trajectories for the different launch angles 
for $\varepsilon = 80 $. The blue line corresponds to  ${\cal C}_m(\varepsilon)$, note that the maximum 
height is $y\sim 0.12$ in contrast to  $y\sim 5.0$ for the drag-free case, however, this can 
be the case of small friction parameter $b$ and large initial velocity $V_0$ giving a large
$\varepsilon$ value. In a black line appears the orbit launched at $\alpha^*$, in red line the corresponding
to attain the maximum range and in blue dashed line the orbit with maximum skewness. 

\section{The synchronous curve}\label{synchronous}

In MacMillan's book\cite{MacMillan} the calculation of the synchronous curve was done for the drag-free case. 
This curve is formed if  many projectiles were fired simultaneously from the same point, each one 
with at different launch angle and same initial speed $V_0$. The locus will be a circle of radius $V_0 t$ and center in 
the point $(0,-\frac{1}{2} g t^2)$, i.e.  
\begin{equation}
x^2+(y+\frac{1}{2} g t^2)^2 = V_0^2 t^2.
\end{equation}
Here, we shall demonstrate that a circle is the synchronous curve in the linear drag case as well.
Following reference \cite{MacMillan}, we eliminate the launch angle $\alpha $ from the position solutions,
in the present case they are equations (\ref{timesolx}) and (\ref{timesol}). We write down $\cos \alpha$ and $\sin \alpha$ and 
rearrange the terms to give,
\begin{eqnarray}
\cos \alpha & = &\frac{b}{V_0}\frac{x}{ \ 1-\exp(-b) \ }, \\
\sin \alpha & = &\frac{b}{V_0} \frac{y-\frac{g}{b^2}(1-bt-\exp(-bt))}{1-\exp(-bt)}.
\end{eqnarray}

Substituting these expressions in the identity $\cos^2 \alpha + \sin^2 \alpha = 1$ we obtain 

\begin{equation}
x^2+(y-y_c(t))^2 = R^2(t).
\end{equation}
With
\begin{equation}
y_c(t) =\frac{g}{b^2}(1-b t - \exp(-b t)),
\label{yc}
\end{equation}
the center and 
\begin{equation}
R(t) = \frac{V_0}{b} (1-\exp(-b t)),
\label{Rc}
\end{equation}
the radius. In order to recover the case where $b \to 0 $, we consider a Taylor expansion for the exponential 
up to second order in the exponential in Eq. (\ref{yc}) and up to first order in Eq. (\ref{Rc}). 
The fact that this circle exists in the presence of a drag force is remarkable.

\section{Conclusions}\label{Conclusions}

We obtained an explicit form for the locus ${\cal C}_m$ composed by the set of maxima  of 
all the trajectories of a projectile launched at an initial velocity $\vec{V}_0$, and in the presence of 
a linear drag force, $-mb \vec{v}$, i.e. ${\cal C}_m$ is the locus of the apexes.
In polar 
coordinates, ${\cal C}_m$ is written in terms of the principal branch of the Lambert $W$ function for 
negative values. This represents the parameterization of the curve by the polar angle $\theta_m$ only
and gives ${\cal C}_m$ in a closed form and exhibits the deep relationship between the Lambert $W$ function
and the linear drag problem. 
The curvature of ${\cal C}_m$ was calculated for different values of the 
dimensionless parameter $\varepsilon \equiv b V_0/g $ in two parameterizations. The first one,
the polar parameterization, shows a maximum that slightly departs from the drag-free 
case in $\theta = 1/2 \arctan(4/3)$. A wider exploration of the functional dependence respect 
to $\varepsilon$ is pending due to numerical accuracy in the calculation of the Lambert $W$ function 
near the limit at $x=-1/e$. In the case of a parameterization using the launch angle $\alpha$ 
there is not such a restriction. In this case, the curvature was calculated for a wide range 
of the parameter $\varepsilon$ yielding maximum at  angle values larger than those corresponding 
to maximum range. Comparison with the maximum skewness angle \cite{Steward2006} was also done
and the difference is larger than the previous one. 
As an addendum, we demostrate that the synchronous curve, in this case, is a circle as in the 
drag-free case.

\section*{Acknowledgments}
This work was supported by  PROMEP 2115/35621. HHS thanks to M. Olivares-Becerril for 
useful discussions and encouragement.  

\appendix
\section{Polar form of an Ellipse with origin at bottom.} \label{appendix1}

Ellipse canonical form or the polar form with the origin considered in one of the focus  are 
standard knowledge. In the present case, however, we require to consider the origin of coordinates 
located in the {\it bottom} of the ellipse, since, in the presence of a drag force, the {\it launching origin} 
is the only invariant point when we change the drag force value. 
To obtain the ellipse form, we  depart from the drag-free
solutions at the locus of the apexes,
\be
x_m = \rho \sin \alpha \cos \alpha,
\label{xmdf}
\ee
and 
\be
y_m = \frac{\rho}{2} \sin^2 \alpha,
\label{ymdf}
\ee
being $\rho \equiv \frac{V_0^2}{g}$. With the help of the trigonometric relations 
$\sin 2\alpha = 2 \sin \alpha \cos \alpha$ and $2\sin^2 = 1 - \cos 2 \alpha$ we transform the 
upper equations into 
\ba
\sin 2 \alpha & = & \frac{2 x_m}{\rho}, \\
\cos 2 \alpha & = & 1 - \frac{4 y_m}{\rho}.
\ea
Taking the squares in both expressions, summing them and arranging terms, we arrive to 
\be
\frac{4 r_m}{\rho} \left( \frac{r_m}{\rho} (1+3 \sin^2 \theta_m) - 2 \sin \theta_m \right) = 0.
\ee
Where we used the polar coordinates $x_m = r_m \cos \theta_m$ and $y_m = r_m \sin \theta_m$. 
The solutions are $r_m=0$ and 
\be
r_m(\theta_m) = \frac{2 \rho \sin \theta_m}{1+3 \sin^2 \theta_m}.
\ee
The second one is the required form for the ellipse. 

\section{Drag-free limit for $\tilde r$} \label{appendix2}

In order to obtain the drag-free limit for the locus ${\cal C}_m$  given in Equation (\ref{W1}),
we note that 
\be 
\tan \theta = (1/2) \tan \alpha
\label{tangent} 
\ee 
and that $f(\alpha) = \sin \alpha(\theta_m) \cos \alpha (\theta_m)$.
The first expression is obtainable from the drag-free solutions Eqs. (\ref{xmdf}) and (\ref{ymdf}), and the 
second is obtained by setting $b \to 0$ in Eq. (\ref{fdtheta}). 

The expansion in a power series of the Lambert $W$ function up to first order is just the identity \cite{Knuth1} 
and hence, 
\be 
\begin{array}{lcl}
\tilde{r}(\theta_m) & = & -\frac{1}{\varepsilon^2 \sin \theta_m} W(-\varepsilon^2 \tan \theta_m  f(\alpha)) \\
  & \approx &  -\frac{1}{\varepsilon^2 \sin \theta_m} (-\varepsilon^2 \tan \theta_m  f(\alpha)) \\.
  & = & \frac{\sin \alpha \cos \alpha }{\cos \theta_m}.  \\
  & = & 2 \sin \theta \frac{\cos^2 \alpha}{\cos^2 \theta_m}.
\end{array}
\ee
Where we used relation (\ref{tangent}) in order to obtain the last line. Using trigonometric  identity 
$\sec^2 \alpha - \tan^2 \alpha = 1$ and Eq. (\ref{tangent}) we obtain that 
\be 
\frac{\cos^2 \theta_m}{\cos^2 \alpha} = 1+3 \sin^2 \theta_m.
\ee
Using this result in the expression of $\tilde r$ we obtain the desired result, Eq. (\ref{ellipse0}).


\begin{thebibliography}{oo}
\bibitem{Comment} In Ref. ~\onlinecite{Steward2006}, the author comment that for $\varepsilon = 1$ we obtain the special 
value $\alpha_{skew} = \sin^{-1}(1/\phi)$ with $\phi = (1+\sqrt{5})/2$, the golden ratio. However, the solution 
that appears in the article  is not longer valid for $\varepsilon = 1$. If the solution corresponds to another real root of the 
equation this special value corresponds to the case when the initial speed is equal to the limit speed 
$b/g$. This makes much more intriguing this fact.
\end{thebibliography}
\end{document}